\documentstyle[aps,prl,floats,twocolumn,epsf]{revtex}

\begin{document}

{\bf Reply to Simon's Comment on
``Evidence for an Anisotropic State of Two-Dimensional
Electrons in High Landau Levels''}

Large magneto-transport anomalies of high mobility two-dimensional
electron systems in the quantum Hall regime have recently been reported
by Lilly {\it et al.}\cite{lilly} and subsequently by Du {\it et
al}\cite{du}. Lilly {\it et al.} find that when the Fermi level is near
half filling of a spin-resolved highly excited Landau level (e.g. at
$\nu=9/2$, 11/2, etc.), the measured resistivity at very low
temperatures ($T<100$ mK) becomes both highly anisotropic and
non-linear. These effects are not seen in the ground ($N=0$) or first
excited ($N=1$) Landau level, but begin abruptly in the $N=2$ level and
remain clearly evident in several higher levels. At Landau level filling
fraction $\nu=9/2$, Lilly {\it et al.} found, using a square sample
geometry, a resistance anisotropy ratio $R_{xx} /R_{yy}$ of around 60.
More recent data, shown in Fig.~1, obtained with a higher mobility
square sample having the same contact configuration as in Ref.~1,
reveals a ratio of around $R_{xx} /R_{yy} \approx 3500$ at $\nu=9/2$.
These results suggest that a new, and widespread, class of correlated
electron states exists in highly excited Landau levels. 

Simon\cite{simon} finds, via a classical calculation, that the apparent
anisotropy observed in square samples can be significantly larger than
the intrinsic anisotropy of the microscopic resistivities 
$\rho_{xx}$ and $\rho_{yy}$.
The origin of this effect is simple: in the presence of anisotropy a
current flowing in the ``easy'' direction, say between two point contacts
at the middles of opposite sides of the square, does not spread out as
much as it does in an isotropic system. This ``channeling'' of the current
reduces the voltage drop observed between contact placed at the corners
of the square. Simon estimates that an intrinsic anisotropy ratio of 7
is sufficient to explain the observed factor of 60. Using his results,
we calculate that an intrinsic resistivity anisotropy ratio of about 23
is needed to reconcile the new data shown in Fig.~1. 

Simon's results do not challenge the basic fact that transport in highly
excited Landau levels becomes extremely anisotropic at very low
temperatures. Without microscopic anisotropy his calculation yields
macroscopic isotropy. Thus, we find no reason to alter our opinion that
the evidence strongly suggests the development of a new correlated
electron phase in high Landau levels. The rapid development of both
anisotropy and non-linearity as the temperature is reduced below 100 mK
and the stark contrast between ground and first excited Landau levels
and the $N \ge 2$ levels are only two of the more striking pieces of this
evidence. 
Other indicators of non-trivial many-electron physics include
the highly structured magnetic field dependence of the resistivity away
from half filling of the third and higher Landau level. As Fig.~1 shows,
this structure ultimately manifests itself as deep zeroes in $R_{xx}$ on
either side of the anisotropic features around half filling. As Lilly
{\it et al.} reported, the Hall resistance in these regions is quantized, but
at the same value as the nearby integer quantized Hall states. This
remarkable re-entrance of the quantized Hall effect again points to
unusual collective behavior. 

\begin{figure}
\begin{center}
\epsfxsize=3.3in
\epsfclipon
\epsffile[72 279 531 648]{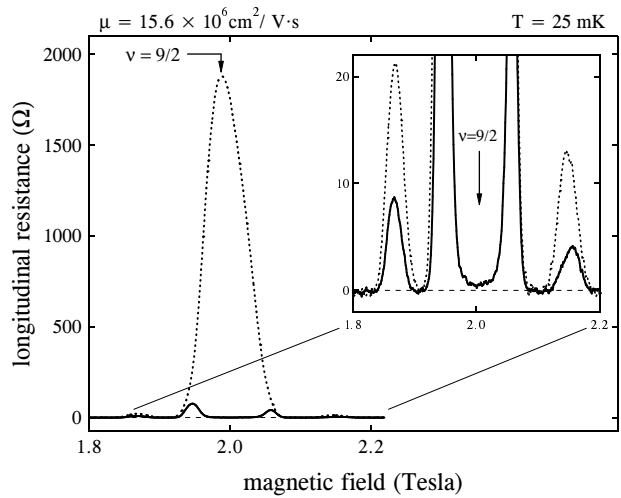}
\end{center}
\caption[figure 1]{Longitudinal resistance at $T=25$ mK around $\nu = 9/2$
for a sample with a very high mobility, $\mu = 15.6 \times 10^6$ cm$^2$/V s,
showing an anisotropy ratio of approximately 3500.  
The two curves represent transport parallel to the $\langle 1 1 0 \rangle$ (solid line)
and $\langle 1 \overline{1} 0 \rangle$ (dotted line) crystal axes.  
Inset: magnified view showing that the resistance minimum at $\nu = 9/2$
in the $\langle 1 1 0 \rangle$ direction remains finite down to 25 mK.
Away from half-filling deep zeroes in the resistance in both directions develop
at 1.90 and 2.10 Tesla.  At these fields the Hall resistance shows integer
quantized Hall plateaus.}
\end{figure}

\vspace{12pt}
\noindent M. P. Lilly, K. B. Cooper, and J. P. Eisenstein\\
California Institute of Technology\\
Pasadena, CA 91125

\vspace{12pt}
\noindent L. N. Pfeiffer and K. W. West\\
Lucent Technologies, Bell Labs\\
Murray Hill, NJ 07974


\begin{references}

\bibitem{lilly}  M. P. Lilly {\it et al.}, Phys. Rev. Lett. {\bf 82}, 394 (1999).

\bibitem{du}  R. R. Du {\it et al.}, Solid State Comm. {\bf 109}, 389 (1999).

\bibitem{simon} S. H. Simon, previous comment (cond-mat/9903086).

\end{references}
\end{document}